# Photothermal tuning of microparticle diffusion anisotropy in nematic lyotropic chromonic liquid crystal

*Patrycja Kadzialka, Lech Sznitko, Pawel Karpinski**

Faculty of Chemistry, Wroclaw University of Science and Technology,

Wyb. Wyspianskiego 27, 50-370 Wroclaw, Poland

e-mail: pawel.karpinski@pwr.edu.pl

**Abstract**

We report a photothermal approach to dynamically tuning the anisotropic diffusion of microparticles suspended in a lyotropic chromonic liquid crystal (LCLC) consisting of a 30 % aqueous solution of Sunset Yellow (SSY). By switching the illumination spectrum between non-absorbing (red) and strongly absorbing (blue) regimes matching the H-aggregate absorption bands of SSY, we demonstrate a significant enhancement in both overall particle mobility and directional diffusion anisotropy ($D_{\parallel}/D_{\perp}$). While off-resonance illumination yields a standard anisotropy ratio of 3.0, resonant blue-light exposure drives a preferential unbinding along the molecular stacks, elevating the anisotropy ratio to 5.5. This light-dependent diffusion anisotropy offers a straightforward framework for optically controlled microfluidic sorting and active soft matter assembly.

## 1. Introduction

Thermally driven Brownian motion of colloidal particles suspended in complex fluids serves as a fundamental probe for non-equilibrium physics, hydrodynamics, and local viscoelasticity.(1,2) In conventional isotropic fluids, such as simple liquids or Newtonian solvents, spatial fluctuations are isotropic, meaning the translational diffusion coefficient $D$ is a scalar quantity governed directly by the classical Stokes-Einstein relationship:(3)

$$D = \frac{k_B T}{6\pi\eta R}, \quad (1)$$

where $k_B$ is the Boltzmann constant, $T$ is absolute temperature, $\eta$ is shear viscosity, and $R$ is the hydrodynamic radius of the particle. (1,2) However, when colloidal particle is embedded in anisotropic liquids – nematic liquid crystals (LCs) – the symmetry of it's diffusion is broken. (1,2,4–6)

In a nematic phase, characterized by long-range orientational order along a space-dependent unit vector called the nematic director $\hat{n}$, the friction experienced by a moving sphere becomes tensor-valued. (1,5–7) Consequently, translational diffusion splits into two distinct modes: parallel ($D_{\parallel}$) and perpendicular ($D_{\perp}$) to the local director $\hat{n}$.(1,2,4–6) Thermotropic liquid crystals, consisting of small organic molecules aligned via dipole-dipole or van der Waals interactions, have historically provided the primary testbeds for investigating anisotropic drag and topological defect structures around inclusions.(1,7,8) Yet, thermotropic systems generally operate within narrow temperature windows, exhibit high sensitivity to organic solvents, and lack environmental compatibility for biological or aqueous applications.(4,9)

Over the past two decades, lyotropic chromonic liquid crystals (LCLCs) have emerged as a unique class of water-soluble liquid crystalline materials. (4,9,10) Unlike conventional lyotropics composed of amphiphilic surfactant molecules that aggregate into micelles due to the hydrophobic effect, LCLCs consist of hydrophilic, disc-like or plank-like aromatic dye molecules, drugs, or nucleic acids.(4,9,11) In aqueous solutions, these flat molecules self-assemble via $\pi - \pi$ stacking interactions into non-covalent, rod-like supramolecular columns (often termed H- or J-aggregates or linear aggregates).(7,11,12) As the dye concentration increases, these supramolecular assemblies elongate and, above a critical volume fraction, spontaneously align to form lyotropic nematic (N) or columnar (M) phases.(4,9,12) The underlying structural units of LCLCs are not fixed rigid rods, but dynamic reversible aggregates whose length distribution depends strongly on concentration, ionic strength, and temperature.(4,9,11,12)

Because LCLC columns are maintained by relatively weak non-covalent interactions ($\sim 10 - 20\ k_B T$ per molecule), their physical properties—such as Frank elastic constants ($K_1$ splay, $K_2$ twist and $K_3$ bend) and Leslie-Ericksen viscosity coefficients—exhibit pronounced temperature dependencies.(4,13) While this structural sensitivity makes LCLCs remarkably versatile for biological templating, sensing and microfluidics,(1,4) controlling local viscoelasticity and colloidal transport in real time without invasive thermal heating stages or chemical dopants remains a major challenge.(1,14)

Photo-induced methods offer exceptional spatiotemporal precision for driving soft matter systems.(14,15) While photo-isomerizable azobenzene dopants or optical tweezers are frequently employed to manipulate liquid crystals,(1,14) direct light-driven alteration of the self-assembled matrix itself represents a powerful alternative.(14–16) In this work, we investigate the two-dimensional translational diffusion of passive polystyrene microparticles embedded in a planarly aligned nematic LCLC matrix composed of a $30$ wt% aqueous solution of the chromonic dye Sunset Yellow FCF (SSY).(4,9) We demonstrate that light resonant with the absorption spectrum of the SSY matrix acts as a localized photothermal driving force.(14–16) This selective excitation alters the dynamic self-assembly of the chromonic stacks, inducing an asymmetric reduction in viscous drag. As a result, particle diffusion along both principal axes is significantly amplified, while the directional anisotropy ratio ($D_{\parallel}/D_{\perp}$) surges from 3.0 to 5.5 without altering the passive, purely Brownian character of the particle trajectories.

## 2. Experimental Setup and Material Characterization

Molecules of SSY self-assemble in water into elongated rod-like stacks via head-to-tail $\pi - \pi$ stacking (H-aggregates),(17) which further align to form a nematic LC phase at high concentrations (Figure 1a,b).(4,9) The local average orientation of these stacks is defined by the nematic director $\hat{n}$.

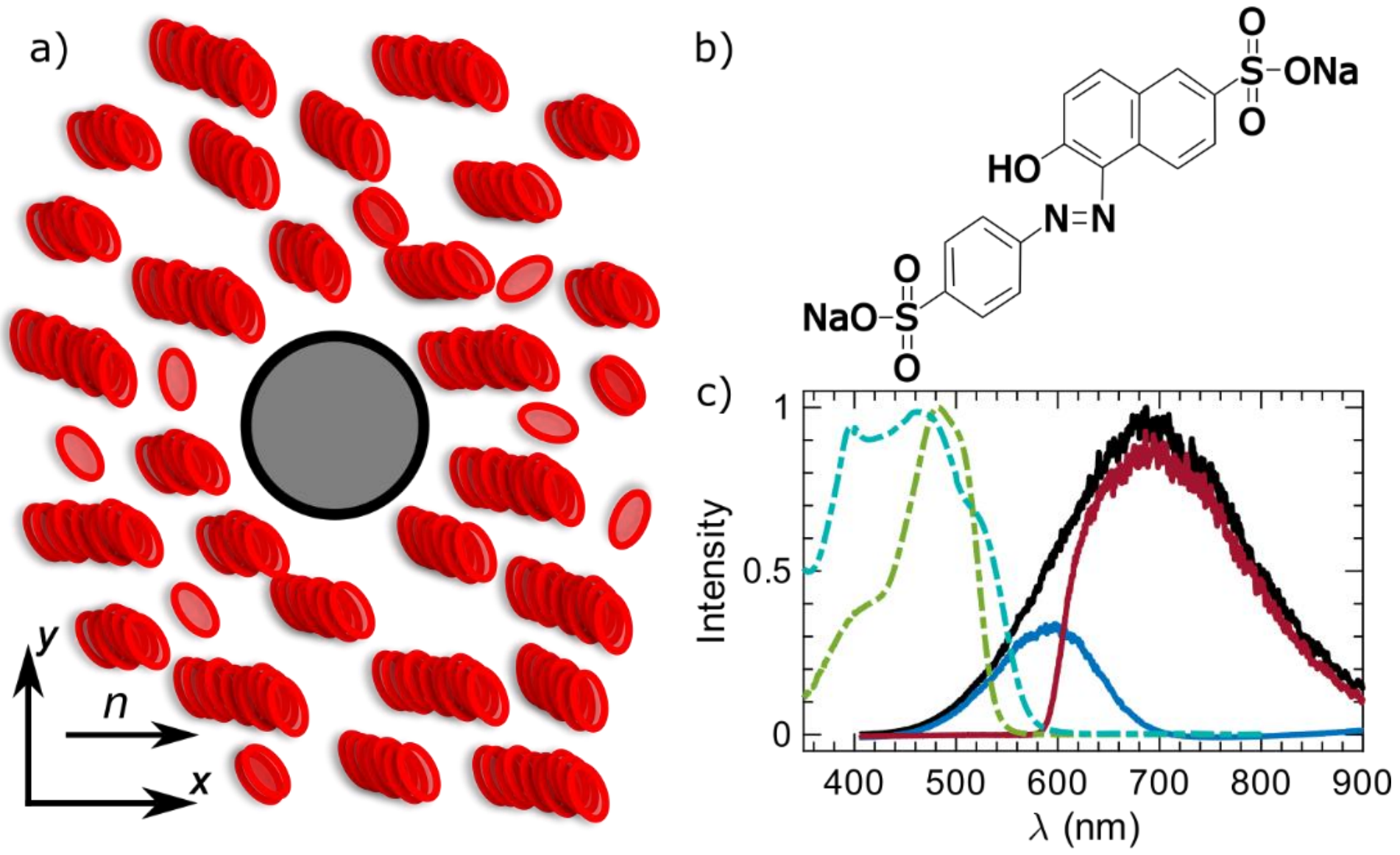

**Figure 1.** a) Schematic illustration of LCLC of SSY, where molecules of SSY (red ellipses) in water are organized in stacks of different lengths, and the rod-like stacks are organized in a higher-order nematic LC structure. The black arrow indicates the director $\hat{n}$ of LC. b) The chemical structure of the SSY molecule. c) Absorption spectra of water solution of SSY dye of concentrations equal to 0.01 and 0.30 $mg/mL$, green and turquoise dashed lines, respectively; and spectra of illumination light, black line original halogen lamp, blue line – blue illumination light, red line – red illumination light.

The spectral characteristics of the system are detailed in Figure 1c. Monomeric SSY (Sigma-Aldrich, Dye content 90 %) exhibits a primary absorption peak around $\lambda \approx 490$ nm (measured at concentration 0.01 $mg/mL$ in twice-distilled water), whereas H-aggregation gives rise to a blue-shifted absorption band near $\lambda \approx 390$ nm (measured at concentration 0.30 $mg/mL$).(12) To probe light-controlled diffusion, we utilized two distinct illumination regimes from a halogen light source: a "blue" excitation spectrum overlapping with the dye's absorption band ($\lambda \approx 400 - 550$ nm), and a "red" off-resonance spectrum ($\lambda > 600$ nm) where SSY absorption is negligible. Polystyrene microparticles of radius $R = 1.25$ µm (microParticles GmbH) were dispersed into the LCLC matrix. The mixture was loaded into a cell of thickness $d = 37$ µm. Planar alignment was achieved by unidirectionally abrading the inner glass substrates with fine abrasive paper (3M WETORDRY 2000). Elastic interactions originating from the quadrupolar or dipolar director distortions force the microparticles to levitate in the bulk near the mid-plane of the cell ($z_{elastic} \approx 10\ \mu m$),(11) suppressing vertical motion ($z$-axis) and restricting the dynamics to pure two-dimensional lateral diffusion. Considering typical LCLC splay, twist and bend elastic constants of the order of: $K_1 = 10\ pN$, $K_2 = 1\ pN$ and $K_3 = 10\ pN$, respectively, what gives an average value of elastic constant $K = 7\ pN$.(4,13) and weak polar surface anchoring $W_a = 10^{-6} - 10^{-5} \frac{J}{m^2}$),(1,18) the de Gennes-Kleman anchoring length is estimated as: $\lambda_{dGK} = \frac{K}{W_a} = 0.1 - 10\ \mu m$. Since $R \sim \lambda_{dGK}$, the microparticles induce relatively weak elastic distortions in the surrounding director field $\hat{n}$, preventing defect-mediated pinning. Particle trajectories were recorded for 10 s at 1000 fps using high-speed video microscopy (microscope: Nikon Ti2, objective: Nikon CFI S Plan Fluor ELWD 60XC NA 0.7, camera: Flir Blackfly S Mono BFS-U3-04S2M, pixel resolution: 115 nm/px) and extracted via particle tracking algorithms (ImageJ MOSAIC). The mean square displacements (MSDs) were calculated across the las time $\tau \in [0.2,1]$ s and averaged over 9000 data points.

## 3. Results and Discussion

Particle trajectories were evaluated along the axes parallel ($x, \parallel$) and perpendicular ($y, \perp$) to the nematic director $\hat{n}$. Figure 2a illustrates representative trajectories recorded under red and blue light exposure.

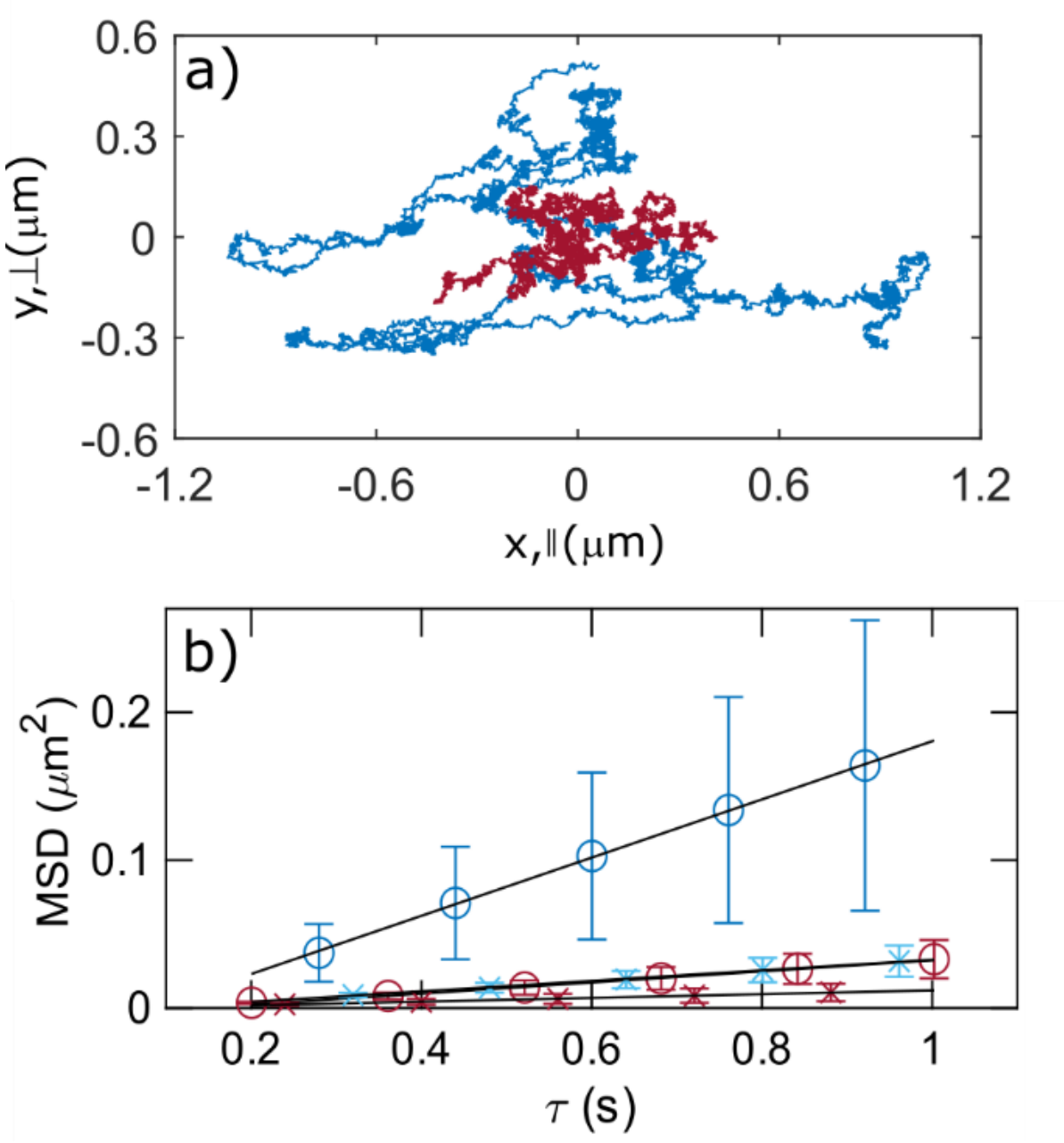

**Figure 2.** a) An example of trajectories of the microparticle illuminated with red (red curve) and blue (blue curve) light. b) MSD of the microparticle displacement in $x$ - parallel to $\hat{n}$ (crosses) and $y$ - perpendicular to $\hat{n}$ (circles) for a microparticle illuminated with red (red markers) and blue (blue markers) light. Markers show only every 160 measurement points for clarity. MSD are averaged over 4 trajectories, error bars are standard deviations. Black lines show linear fits with eq. 2.

MSD as a function of lag time $\tau$ were fitted using the linear 2D diffusion relation:

$$MSD(\tau) = 4D\tau + C. \tag{2}$$

To eliminate localization errors and transient elastic cage effects, linear fits were evaluated for $\tau > 0.2\ s$, see Figure 2b. The extracted diffusion coefficients are summarized in Table 1.

**Table 1.** Anisotropic diffusion coefficients ($D_{\parallel}, D_{\perp}$) and anisotropy ratios ($D_{\parallel}/D_{\perp}$) under red and blue illumination.

| Illumination Mode | $D_{\parallel}\left(10^{-3}\frac{\mu m^2}{s}\right)$ | $D_{\perp}\left(10^{-3}\frac{\mu m^2}{s}\right)$ | $D_{\parallel}/D_{\perp}$ |
|---|---|---|---|
| Red | 9.4 | 3.2 | 3.0 |
| Blue | 49.2 | 8.9 | 5.5 |
| Ratio (Blue/Red) | 5.2 | 2.8 | |

Under red-light illumination, the system displays conventional anisotropic Brownian motion with $D_{\parallel} = 9.4 \times 10^{-3}\frac{\mu m^2}{s}$ and $D_{\perp} = 3.2 \times 10^{-3}\frac{\mu m^2}{s}$, yielding an anisotropy ratio of $D_{\parallel}/D_{\perp} \approx 3.0$. This value aligns well with the Stokes drag predictions for nematic fluids with planar director alignment. (2,7) To cross-check these values against theoretical estimates, the isotropic diffusion coefficient $D_{SSY}$ calculated via the Stokes-Einstein formula eq. 1 using the macroscopically measured bulk viscosity ($\eta_{SSY} \approx 35.2 \pm 11 \times 10^{-3}\frac{kg}{m\,s}$ at 21 °C,measured using IKA Rotavisc) yields: $D_{SSY} \approx 5 \times 10^{-3}\frac{\mu m^2}{s}$, which falls cleanly between the measured parallel and perpendicular values.

In a nematic liquid crystal, the anisotropic drag experienced by a sphere is dictated by the anisotropic viscosity tensor, which can be described via the Miesowicz viscosities ($\eta_a$, $\eta_b$ and $\eta_c$).(2,5,6) For motion parallel to the director $\hat{n}$, the fluid flow primarily involves shear along the molecular alignment, governed predominantly by the lower viscosity coefficient $\eta_b$ and leading to higher values of $D_{\parallel}$.(2,6) Motion perpendicular to $\hat{n}$ requires the sphere to displace and bend adjacent nematic columns, bringing the higher viscosity coefficients $\eta_a$ and $\eta_c$ into play, causing lower values of $D_{\perp}$.(2,6)

When the illumination is switched to blue light ($\lambda \approx 400 - 550$ nm), which strongly overlaps with the absorption band of SSY,(12) particle mobility undergoes a profound transformation. Rather than exhibiting ballistic motion or active self-propulsion (such as that observed in chemically active swimmers or Janus colloids),(19–24) the trajectories remain purely stochastic, linear in MSD vs. $\tau$, confirming that the dynamics remain passive and thermal in nature. (1,2) However, the magnitude of diffusion along both directions increases substantially: $D_{\parallel}$ increases by a factor of 5.2, whereas $D_{\perp}$ increases by 2.8. Consequently, the diffusion anisotropy ratio $D_{\parallel}/D_{\perp}$ surges from 3.0 to 5.5.

This marked enhancement in both overall diffusion rates and directional anisotropy cannot be explained by a simple uniform increase in isotropic temperature.(14) If the system experienced straightforward global Joule heating, the temperature $T$ rise would cause a modest, nearly isotropic decrease in effective fluid viscosity, leading to proportional increases in both $D_{\parallel}$ and $D_{\perp}$, thereby preserving or decreasing the anisotropy ratio $D_{\parallel}/D_{\perp}$.(2,14) The increase of $D_{\parallel}/D_{\perp}$ indicates a fundamental, structural change in the viscoelastic properties of the chromonic matrix under resonant illumination.(4,9,14)

We speculate that the underlying physical mechanism relies on localized photothermal softening of the chromonic supramolecular assembly.(12,14) SSY H-aggregates act as efficient nano-absorbers, and upon absorbing blue light photons, the molecules undergo non-radiative decay, converting optical energy directly into local thermal energy within the chromonic stacks. Because LCLC aggregates are held together by weak non-covalent $\pi - \pi$ interactions, localized heating disrupts the physical cross-linking and stack elongation.(4,9,12) According to supramolecular assembly models for LCLCs, raising the local micro-temperature causes the average column length $\langle L \rangle$ to decrease rapidly:(4,9,12)

$$\langle L \rangle \sim \exp\left(\frac{E_{\mathrm{sci}}}{2k_{\mathrm{B}}\mathrm{T}}\right), \qquad (3)$$

where $E_{\mathrm{sci}}$ is the scission energy required to break an H-aggregate stack [3,5]. A reduction in column length $\langle L \rangle$ directly lowers the local nematic order parameter $S$ and dramatically reduces the effective viscosity along the director $\hat{n}$.(4)

Importantly, this photothermal column scission impacts parallel and transverse hydrodynamics in an inherently asymmetric manner.(2,25) Sliding drag along the director $\hat{n}$ is governed by intermolecular friction between adjacent chromonic columns. (4) As columns break into shorter segments, longitudinal sliding resistance drops precipitously, allowing microparticles to diffuse with significantly less friction along $x, \parallel$.(4) On the other hand, motion perpendicular to $\hat{n}$ $(y, \perp)$ requires the particle to distort and displace the surrounding nematic continuum, which is constrained by the Frank elastic constants ($K_1$ splay and $K_3$ bend).(2,25) While elastic constants also weaken slightly with temperature, the elastic barrier to transverse displacement remains substantial compared to the dramatic reduction in longitudinal shear viscosity.(2,4,25)

This preferential unbinding along the molecular stacks selectively "softens" the medium parallel to the director field, (4) driving the observed amplification of the anisotropy ratio from 3.0 to 5.5. Similar anisotropic photothermal responses have been suggested in liquid crystalline networks and complex soft matter,(12,14) but its direct realization as a means to dynamically tune passive colloidal transport in aqueous chromonic systems highlights the exceptional sensitivity of LCLCs to optical fields.

## 4. Conclusions

In conclusion, we have demonstrated a simple, non-invasive, all-optical approach to dynamically modulate the anisotropic diffusion of passive microparticles suspended in a lyotropic chromonic liquid crystal. By switching between off-resonance (red) and resonant (blue) illumination matching the absorption band of Sunset Yellow H-aggregates, we exploit localized photothermal conversion to alter the supramolecular structure of the host matrix. (9,12,14)

Resonant blue light absorption induces localized heating, breaking the non-covalent $\pi - \pi$ stacks and shortening the average chromonic column length $\langle L \rangle$ (eq. 3).(4,9,12) This structural softening selectively lowers the longitudinal shear viscosity parallel to the nematic director $\hat{n}$, while transverse elastic resistance remains higher.(2,13,25) As a consequence, both diffusion coefficients $(D_{\parallel}, D_{\perp})$ increase, but the directional anisotropy ratio $D_{\parallel}/D_{\perp}$ is elevated from 3.0 to 5.5. Throughout this process, the particle motion remains strictly passive and Brownian, governed by thermal fluctuations within an optically modified viscoelastic medium.

This light-responsive transport mechanism requires no chemical functionalization of the particles or complex synthetic photo-switchable molecules. Looking forward, the ability to optically tune transport anisotropy in biocompatible LCLCs opens promising avenues for microfluidic particle sorting, directional transport in lab-on-a-chip devices, optically guided colloidal self-assembly, and adaptive soft-robotic systems.

**Acknowledgement**

We would like to acknowledge the National Science Center project SONATA no.: 2019/35/D/ST3/02272 for financial support.

**Data Availability Statement**

The data that support the findings of this study are openly available in RepOD at 10.18150/QUYQ0I.